# Extremely Weakly Interacting $\Delta S_z = 0$ and $\Delta S_z = 1$ Excitations and Evidence for Fractional Quantization in a Magnetization Plateau: CeSb

P. G. LaBarre[1], B. Kuthanazhi[2,3], C. Abel[2,3], P. C. Canfield[2,3], and A. P. Ramirez[1]

[1] Department of Physics, University of California Santa Cruz, Santa Cruz, California 95064 USA
[2] Department of Physics and Astronomy, Iowa State University, Ames, Iowa 50011, USA
[3] Ames Laboratory, Iowa State University, Ames, Iowa, 50011, USA

*Abstract:*

The plateau at 1/3 of the saturation magnetization, $M_s$, in the metamagnet CeSb is accompanied by a state of ferromagnetic layers of spins in an up-up-down sequence. We measured $M$ and the specific heat, $C$, in the plateau, spin wave analyses of which reveal two distinct branches of excitations. Those with $\Delta S_z = 1$ as measured by $M$, coexist with a much larger population of $\Delta S_z = 0$ excitations measured by $C$ but invisible to $M$. The large density of $\Delta S_z = 0$ excitations, their energy gap, and their seeming lack of interaction with $\Delta S_z = 1$ excitations suggest an analogy with astrophysical dark matter. Additionally, in the middle of the plateau three sharp jumps in $M(H)$ are seen, the size of which, $0.15 \% M_s$, is consistent with fractional quantization of magnetization-per-site in the down-spin layers.

Over the last few decades, magnetization plateaus have provided a rich source for the study of quantum magnetism. These plateaus have been shown to arise in metamagnets, low-dimensional magnets, and geometrically frustrated systems and are understood to result from a gap in the many-body energy spectrum analogous to that underlying the *fractional* quantum Hall effect (QHE) [1,2]. In such systems, a physical quantity is quantized over specific magnetic field ranges due to a topological commensurability between localized excitations and the lattice, resulting in translational symmetry breaking [3]. For spin models, this quantization condition is expressed as $n(S-m)$ = integer in the plateaus, where $n$ is the translational period of the ground state, $S$ is the total spin per site, and $m$ is the magnetization per site in units of spin quantum number [1,3].

In order to better understand topological quantization in magnetic systems, we chose the metallic metamagnet CeSb, which in zero field ($H$) undergoes simultaneous antiferromagnetic ordering and a structural transition from its high temperature NaCl phase to tetragonal symmetry at $T_N$ = 16 K [4]. At finite $H$ and lower $T$, CeSb exhibits a large number of ordered phases which have been identified and extensively studied by magnetization [4-9], specific heat [4,10-13], neutron diffraction [6-8,12,14-19], x-ray diffraction [19,20], ARPES [21] and charge transport [9,22,23]. The $Ce^{3+}$ moments in CeSb arise from a J = 5/2 ($\Gamma_7$) doublet ground state (equivalent to an effective $S$ = 1/2 and an effective $g$-factor $g = 4.28$) and an anisotropic spin-spin interaction, attributed to strong mixing of the Ce *f*-states with neighboring *p*-orbitals in Sb [5,18]. While the magnetic response is anisotropic with an easy axis [5], inelastic neutron scattering sees dispersive magnetic excitations [14,15] indicating only partial Ising character. In the ordered states, the spins are ferromagnetically aligned in layers perpendicular to (0,0,1) with the multiplicity of states corresponding to different repeat patterns of spin planes, aligned and anti-aligned with $H \parallel (0,0,1)$ [12]. The plateau corresponding to 1/3rd of the saturation magnetization, $M_s$, occurring for $2.1 \leq H \leq 3.7$ T and $T < 10$K, is defined by a quantization condition among layers which triples the unit cell in an up-up-down (↑↑↓) sequence [1,3], and can be relatively flat in high quality crystals [9]. The quantization condition parameters for this plateau are thus $n = 3$, $S = 1/2$, and $m = (1/3)(1/2) = 1/6$.

In this work we present studies of the magnetization, $M(T, H)$, and specific heat, $C(T, H)$, in the 1/3rd plateau ($n = 3$) of CeSb. We find that the slope of *M v H* or "plateau susceptibility", $\chi_P$, in the 1/3rd phase is $1.15 \times 10^{-2}$ emu/mole and weakly temperature dependent between 2-9 K. Application of spin wave theory to explain the temperature dependences in $M$ and $C$ leads to a large discrepancy in the spin wave stiffness between these two quantities, indicating the presence of two distinct excitation branches. The $\Delta S_z = 0$ excitations probed by $C$ are likely due to time-reversed pairs of spins across interfaces between up and down layers, revealing a spin flip density in the down layers of 9% at the high-temperature boundary of the plateau. The seeming invisibility of the large population of $\Delta S_z = 0$ excitations to $M$ suggests an analogy to dark matter. Furthermore, on close inspection, the shape of the plateau resembles a Brillouin function, characteristic of free spins, on top of which we observe distinct jumps whose magnitude and width suggest two-dimensional long-range ordering among $\Delta S_z = 1$ excitations in the down layers. Moreover, by subtracting an arbitrary free-spin background, the jumps reveal plateaus which are consistent with fractional quantization condition *within* the down layers.

Crystals of CeSb were grown from a flux (Supplemental Material) and have a cubic morphology with an average cube side of 5 mm and mass of approximately 60 mg. For this work, we confirmed reproducibility of the data shown below on three different samples from two growth runs. The dc magnetization was measured in a Quantum Design superconducting quantum interference device (SQUID) magnetometer with hysteresis protocols described below for $H$ along a cubic axis to an accuracy of ~1 degree. Measurements of $C(T)$ were made using the relaxation method in a Quantum Design Physical Property Measurement System (PPMS). The lattice contribution to $C(T)$ in CeSb was identified with the specific heat of a single crystal of non-magnetic LaSb, grown for this purpose. All data presented here have this lattice contribution subtracted, and thus represents the magnetic contribution to $C(H,T)$ see Supplemental Material for lattice subtraction methods, which includes Refs. [24-26].

In figure 1(a), we present *M(H,T)* for both increasing and decreasing external fields, and find consistency with previous results. We note that the temperature dependence of $M(H)$ within the plateau is *i)* small, with $dlnM/dT \cong 0.0013 K^{-1}$ and, *ii)* much less for decreasing $H$ than for increasing $H$, a point to which we will return later. As $H$ increases out of the plateau and into the aligned paramagnetic phase, $M$ reaches its saturation value $M_s = 1.116 \times 10^4$ emu/mole = $2.0\mu_B$, as observed previously [9]. The deviation from the free-ion value of 2.14 $\mu_B$ is ascribable to the presence of the near-lying $\Gamma_8$ quartet [27]. In figure 1(b), we present $C(H)/T$ from $H = 1 - 5$T, taken on increasing field with the 1/3rd plateau clearly delineated as a mesa-like feature indicating an increase in available states within the plateau. Most importantly, the temperature dependence of $C(H)/T$ taken between 2K and 9K is $dln(C/T)/dT \cong 0.29 K^{-1}$, i.e. a factor of 200 times greater than $dlnM/dT$. By contrast, usual antiferromagnets such as $Dy_3Al_5O_{19}$ display $dln(C/T)/dT \approx dlnM/dT$ [28].

Before addressing the sub-gap structure in $M(H)$, we attempt to model the dramatically different temperature dependences of $M$ and $C/T$ by spin waves. Viewing the FM layers as independent systems, as $H$ is increased the up layers see an *increasing* local field and the down layers see a *decreasing* local field. This dictates a decreasing spin wave density with $H$ for the up layers (larger local gap) and an increasing spin wave density for the down layers (smaller local gap). Thus, in the independent-layer scenario, the combined effect implies that the down layer will dominate temperature-induced changes in $M$. Following Niira's modification [29] of Bloch's theory to include an anisotropy-induced spin wave gap, $\Delta$, the temperature dependence of $M$ can be expressed as $M(T)/M_s = 1 - AT^{3/2}e^{-\Delta_M/T}$, where $A$ depends on fundamental constants, $M_s$, and the spin wave stiffness, and the gap $\Delta_M$ depends on the internal magnetic field. As shown in figure 2 (a)/(b), we find good agreement to spin wave fits of $M(T)$ and a corresponding form for $C(T)/T$ [30] at constant fields in the 1/3$^{rd}$ phase. For the fits to $M(T)$ and $C(T)$ we used the data obtained on increasing $H$ and we discuss the down sweeps separately below. The spin wave stiffness for $M(T)$ and $C(T)$ was found to be 111-119 $K\text{Å}^2$ and 10.3 $K\text{Å}^2$ respectively. The spin wave gaps (figure 2 (c)) extracted from the fits of $M$ and $C$ are $\Delta_M = 15 - 20K$ and $\Delta_C = 22.9\,K$, respectively. In contrast to the large difference in spin wave stiffness, the gap values from $C$ and $M$ are within 50% of each other, showing that the two excitation types probed by $C$ and $M$ see similar gaps but with vastly different density of states. The integrated entropy $\Delta S = \int C/T$ from 2K to 9K in the plateau is 9%Rln2.

The simple spin wave analysis above belies the complex microscopic nature of the charge-spin-orbital degrees of freedom at play in CeSb. Although the degeneracy-lifting represented by the various magnetic transitions below $T_N$ is among J = 5/2 doublets, the energy gap to the excited S = 3/2 ($\Gamma_8$) quartet is only 24K [13], i.e. comparable to the gap inducing the plateau. Such an excited state for free spins in a field of 3 T would produce a total entropy between 2-9K, (1/3)S$_{quartet}$ = 2.36 J/mole K, greatly exceeding that of CeSb, S = 0.55 J/mole K (See Fig. 2, Supplemental Material). Thus, because the density of states for these crystal field excitations is significantly reduced compared to that of free spins, the effect of the energy gap associated with 1/3 quantization extends to the suppression of the excited state population. Ambiguity in the precise nature of the $\Delta S = 0, 1$ excitations does not alter our main observation, however, that they are not mutually interacting.

The situation of two types of excitations equal in energy but not interacting is reminiscent of Herring's description of accidental degeneracy of free electrons [31]. In magnetic systems, similarity is found in frustrated s = ½ Heisenberg spin models such as the frustrated spin chain [32,33] and the antiferromagnetic kagome lattice [34], both of which exhibit magnetization plateaus. For the kagome lattice at $H = 0$, however, both exact diagonalization and mean field analyses reveal distinct sectors of gapped triplet excitations coexisting with gapless singlet excitations [35,36]. The present situation differs from these $T = 0$ analogues in that the $\Delta S_z = 0$ excitations possess a gap reflecting the many-body spectrum and thus would arise from a mass term in an effective Hamiltonian. That the $\Delta S_z = 0$ excitations are simultaneously dense in the down layer while also invisible to $M(H)$ suggests an analogy to astrophysical dark matter. While the origin of dark matter is of course unknown, the present situation is more tractable. Since the mini jumps are likely due to ordering among localized $\Delta S_z = 1$ excitations, an excitation not resulting in a change of $M$ must involve flips of two spins, one aligned and one anti-aligned with $H$. Thus, one of the two spins must come from the down-layer and the other from a nearby up-layer. If half of the 9.0% Rln2 entropy developed up to 9K is in the down layer, which represents only 1/3rd of all spins, then approximately 15% of the down layer spins will be reversed. This large density of spin flips should lead to a similar effect in the mean field, evinced in both the plateau as well as $\Delta_C$, as shown in Fig. 2c.

The astrophysical analogy suggested above can be extended by considering that the term "dark" corresponds to the absence of elastic scattering of photons from a large amount of mass inferred from observations of galaxy expansion. One might ask whether an analogue of photons, namely neutrons, are able to image $\Delta S_z = 0$ excitations in CeSb. While it is unlikely that such excitations are directly observable in the neutron elastic channel, their presence might be inferred from accurate diffraction measurements as a decrease in ordered moment of the ↑↑↓ phase as $T$ is increased within the plateau. Of course the present system also admits the possibility of *inelastic* scattering, which should allow direct imaging of $\Delta S_z = 0$ excitations as a flat band at the gap energy. Such a measurement would complement the specific heat by providing a way to image magnetically-dark excitations using inelastic radiative processes. A class of systems with more relevance to CeSb are the QHE systems. Despite evidence that the fractional ground states are spin polarized, thus lacking a singlet degree of freedom, the observation of oscillations in $C(H)$ measurements [37] makes it reasonable to ask if measurements of the $T$-dependence would reveal magnetically-dark excitations.

We now address the mini jumps in the middle of the plateau. Above a Brillouin-function-like background, three mini jumps are seen at 2.7, 2.9, and 3.1 T for $H$ up-sweeps and one distinct jump at 2.85 T for $H$ down-sweeps, as shown in the inset of figure 1(a). Often, small jumps in $M(H)$ are observed close to the major jumps straddling a plateau and are associated with incomplete ordering related to sample inhomogeneity. By contrast, the mini-jumps in CeSb occur in the middle of the 1/3$^{rd}$ plateau. In addition, we observe these jumps in three different samples at virtually the same field values, as shown in Fig. 4 (b), which argues against an extrinsic explanation. The jumps are characterized by their size, ~0.0015 of $M_s$, and by their width, which is comparable to the width of the major jumps. No corresponding features are observed in $C(H)$, consistent with the lower precision of this measurement compared to SQUID-based $M(H)$.

Since the large jumps in $M(H)$ occur at plateau boundaries, it is reasonable to ask if the mini-jumps are associated with mini-plateaus, and thus with an additional topological quantization condition. As discussed above, such a quantization condition relates the jump height inversely to the period of broken spatial symmetry. If the broken symmetry is in the same direction as that associated with the 1/3$^{rd}$ quantization, (001), then $0.0015 M_s$ implies a period of ~670 lattice constants, which is extremely unlikely. More likely is that the spatially broken symmetry is transverse to (001). Moreover, due to the ↑↑↓ order within the plateau, it is reasonable to assume that the mini-jumps correspond to spin-flip ordering only within the ↓-layer since the ↑-layer spins should be nearly fully polarized along $H$. In this case, for the largest of the mini-jumps such spin flips would be separated by $((1/3 M_s)/0.0015 M_s)^{1/2} \sim 15$ lattice constants in the down layer and the smaller "satellite" jumps indicate a larger separation. While such a repeat period is still quite large on the scale of usual atomic spin ordered wavelengths, we note that Skyrmion lattices can exhibit such large lattice constants [38].

If, as argued above, the mini-jumps correspond to plateaus related to spatial symmetry breaking in the ↓-layers, then the finite $M(H)$ slope in the 1/3$^{rd}$ plateau masks the mini-plateaus. Is it possible, then, to separate the spin-wave background from the mini-plateau $M(H)$? Since the spin wave analysis above describes $M(H)$ well and since it has the shape of a Brillouin function, B(x), we model the background spin-wave contribution as $M_{\text{Brillouin}} = ngS\mu_B B(x)$ where $n$ is a molar spin density, $x = gS\mu_B H_{P,\text{eff}}/k_B T$, and $H_{P,\text{eff}} = H - H_{P,0}$, with $H_{P,0} = 2.4 - 2.55\ T$, from 9-2 $K$ respectively, being the lower boundary at which the Brillouin function fit was made. We subtract from each up-sweep isotherm $M_{\text{Brillouin}}(H,T)$, with $n$ adjusted to achieve flatness in $M - M_{\text{Brillouin}}$. As shown in figure 3, for each isotherm a single B($x$) function produces plateau-like regions for all three mini jumps and the inset of figure 3 shows the values of $n$ required to produce such flatness. Since the spin-wave spectrum is probably quite complex, given the different sublattices and exchange constants [18], it is surprising that a simple Brillouin function parameterization produces such flat mini-plateaus.

For any plateau system, the magnitude of field where $M(H)$ enters a plateau is significant only for providing information on the internal mean field. Nevertheless, it is of possible future interest to note the precise location of the mini-jumps. In figure 4(a), $dM/dH$ for the jumps provides their precise positions and variation with $H$ and $T$. We see that the central, and largest, of the jumps asymptotically approaches as $T \to 2K$, a field value corresponding to $H = H_l + f_p \Delta H_{1/3}$ where $\Delta H_{1/3} = 1.7 \pm 0.1\ T$ is the width of the 1/3$^{\text{rd}}$ plateau and $f_p = 1/2$ (see Supplemental Material). Small variations in the width, lower, and upper field boundaries are seen from $2 - 9\ K$, which adjust the $f_p = 1/2$ field value at each temperature. The $f_p = 1/2$ field values appear to converge upon decreasing $T$ (open circles Fig. 4 (a)). The other two jumps seem to approach $f_p = 3/8$ and 5/8.

Finally, we address the hysteresis of $M(H)$ observed in the 1/3$^{\text{rd}}$ plateau region. Such hysteresis is not uncommon and usually appears near the steps straddling the plateau. The cause of hysteresis in such cases is likely similar to hysteresis in $M(H)$ loops of hard ferromagnets, namely pinning of domain walls by defects or impurities. While such a mechanism may explain the small difference in the magnitude of $M(H)$ between up and down sweeps, $\Delta M_{\text{up-down}} \approx 0.5\%$, it doesn't immediately explain why only the $f_p = 1/2$ mini-jump is seen on down sweeps. The larger $M(H)$ for down-sweeps than for up-sweeps implies, however, that a residual density of spins in the down layer are pinned in the up direction. The effect of such a defect density can be seen in the broadening of the $f_p = 1/2$ mini-jump on down-sweeps which is likely due to an average defect-defect distance comparable to 15 lattice constants. A distance of this magnitude could then eliminate the satellite mini-jumps since it would correspond to a density greater than the 2D percolation threshold of those fractionalized spins.

To summarize, we have shown via $C(T)$ measurements within the 1/3$^{\text{rd}}$ plateau of CeSb, the presence of a separate $\Delta S_z = 0$ branch of excitations invisible to $M(H)$ and interacting with $\Delta S_z = 1$ in a manner that evokes dark matter in the astrophysical context. In addition, the $\Delta S_z = 1$ excitations reveal sharp small jumps in $M(H)$ that imply additional translational symmetry breaking in the down-spin layers. Further study of these phenomena will include small angle neutron scattering, ultra-low temperature magneto-thermal and transport measurements, and theoretical studies. Also of interest would be an exploration of the integer and fractional QHEs to test for excitations analogous to the $\Delta S_z = 0$ magnetically-dark excitations we have observed in CeSb.

Acknowledgements: We would like to thank Sriram Shastry, who suggested the dark matter analogy, Gabriel Aeppli, Leon Balents, Daniel Cox, Jim Eisenstein, Morten Eskildsen, Roderich Moessner, Mohit Randeria, and Peter Schiffer for valuable conversations. The measurements were performed at UCSC and supported by U. S. Department of Energy Office of Basic Energy Science, Division of Condensed Matter Physics grant DE-SC0017862. Growth and basic characterization of CeSb and LaSb single crystals were performed at Iowa State University Department of Physics and Ames Laboratory and was supported by U. S. Department of Energy, Office of Basic Energy Science, Division of Materials Sciences and Engineering under Contract No. DE-AC02-07CH11358.

Figures:

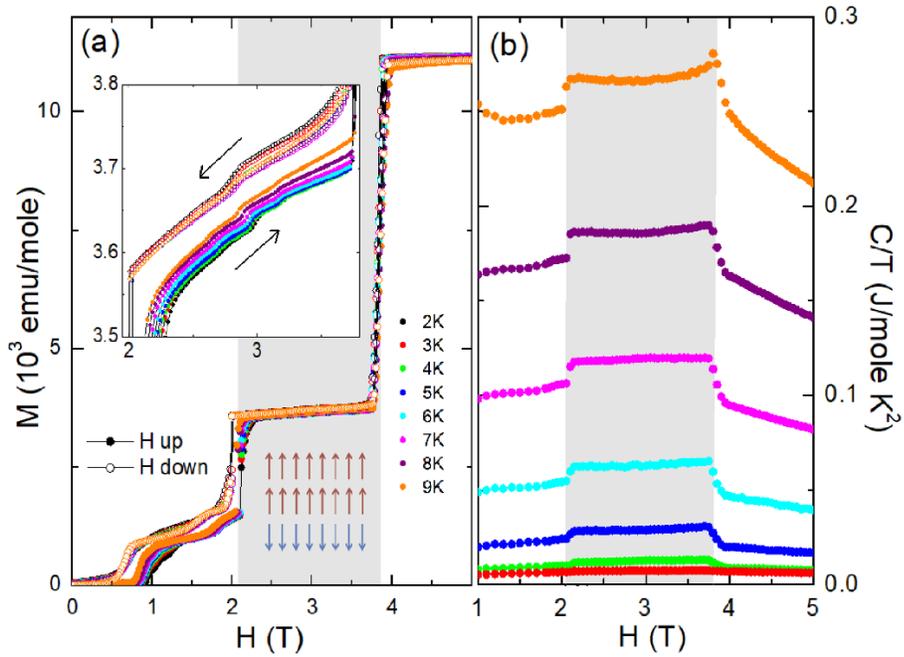

Figure 1: (a) *M v H* at different temperatures between 2 and 9 K. Solid circles are data points taken upon increasing magnetic field and open circles are data points taken upon decreasing magnetic field. Inset- zoomed in plot of *M v H* within the 1/3$^{rd}$ plateau. (b) *C/T v H* at different temperatures between 2 and 9 K with the grey area denoting the 1/3$^{rd}$ plateau region.

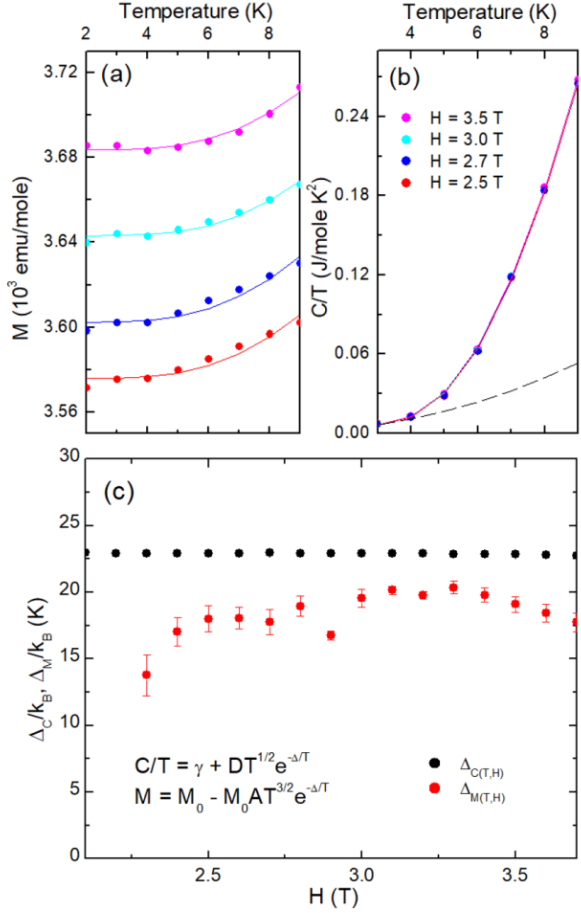

Figure 2: (a) Magnetization vs. temperature at fixed fields, the values of which are given in frame (b). The solid lines are fits to a spin wave form. (b) Specific heat vs. temperature at different fields (data lie on top of each other). The solid lines are fits to a spin wave form and the dashed line obeys $C/T = a + bT^2$, showing that phonons cannot produce the observed temperature dependence. (c) The energy gaps resulting from the spin wave fits for both magnetization and specific heat. For all fits, $\gamma = 5$ mJ/moleK$^2$, $D = 1.11$ J/moleK$^{5/2}$ and $A = -0.0023\ K^{-3/2}$ are held constant.

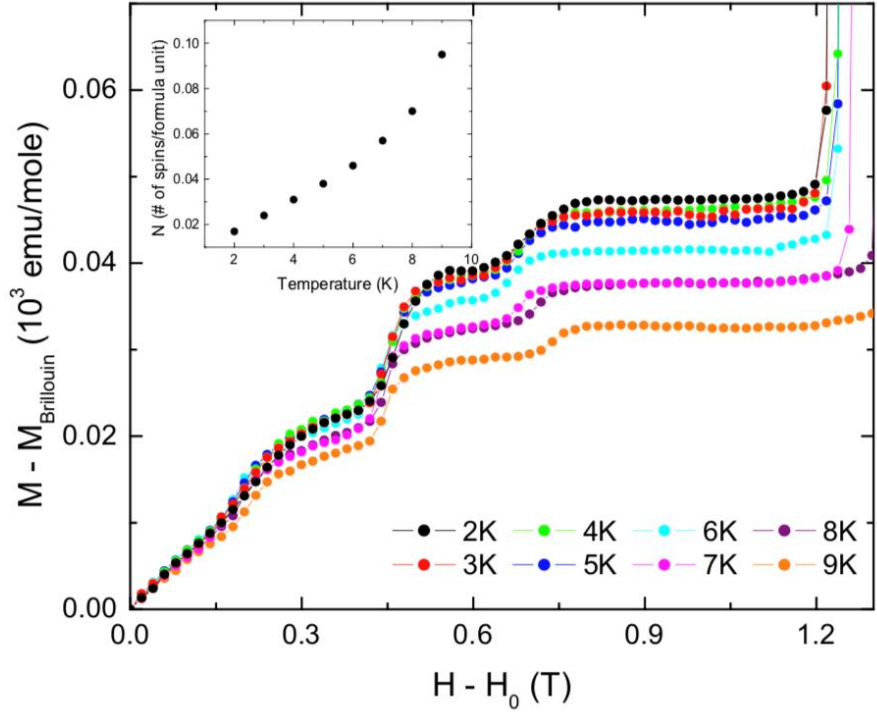

Figure 3: Magnetization data on field upsweep in the 1/3rd plateau vs. magnetic field difference from the lower boundary, $H_{P,0}$, of the plateau region. The data shown are those in Fig. 1 with a Brillouin function subtracted, as described in the text. The inset shows the fraction of free spin density per formula unit used in the Brillouin function subtraction.

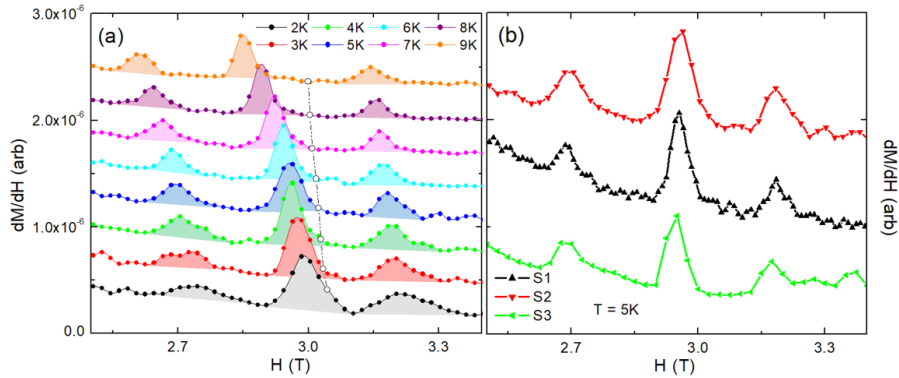

Figure 4: (a) Field derivative of the magnetization in the 1/3rd plateau region vs. field, indicating the positions of the mini jumps. The open circles mark the $f_p = 1/2$ field value for each temperature. (b) Positions of the mini jumps in multiple samples of CeSb. Sample peaks were scaled to comparable magnitudes and set relative to the field width of the 1/3$^{rd}$ plateau in each sample.

References:

[1]     M. Oshikawa, M. Yamanaka, and I. Affleck, Physical Review Letters **78**, 1984 (1997).
[2]     W. Shiramura, K. Takatsu, B. Kurniawan, H. Tanaka, H. Uekusa, Y. Ohashi, K. Takizawa, H. Mitamura, and T. Goto, Journal of the Physical Society of Japan **67**, 1548 (1998).
[3]     M. Oshikawa, Physical Review Letters **84**, 1535 (2000).
[4]     F. Hulliger, M. Landolt, H. R. Ott, and R. Schmelczer, Journal of Low Temperature Physics **20**, 269 (1975).
[5]     G. Busch and O. Vogt, Physics Letters A **A 25**, 449 (1967).
[6]     J. Rossat-mignod, P. Burlet, J. Villain, H. Bartholin, W. Tchengsi, D. Florence, and O. Vogt, Physical Review B **16**, 440 (1977).
[7]     J. Rossat-mignod, P. Burlet, H. Bartholin, J. Villain, D. Florence, and O. Vogt, Physica B & C **86**, 129 (1977).
[8]     J. Rossat-mignod, J. M. Effantin, P. Burlet, T. Chattopadhyay, L. P. Regnault, H. Bartholin, C. Vettier, O. Vogt, D. Ravot, and J. C. Achart, Journal of Magnetism and Magnetic Materials **52**, 111 (1985).
[9]     T. A. Wiener and P. C. Canfield, Journal of Alloys and Compounds **303**, 505 (2000).
[10]    J. Rossat-mignod, P. Burlet, H. Bartholin, O. Vogt, and R. Lagnier, Journal of Physics C-Solid State Physics **13**, 6381 (1980).
[11]    T. Kasuya, Y. S. Kwon, T. Suzuki, K. Nakanishi, F. Ishiyama, and K. Takegahara, Journal of Magnetism and Magnetic Materials **90-1**, 389 (1990).
[12]    J. Rossat-mignod, P. Burlet, S. Quezel, and O. Vogt, Physica B & C **102**, 237 (1980).
[13]    G. Busch, W. Stutius, and O. Vogt, Journal of Applied Physics **42**, 1493 (1971).
[14]    A. Furrer, W. Halg, and H. Heer, Journal of Applied Physics **50**, 2040 (1979).
[15]    G. Meier, P. Fischer, W. Halg, B. Lebech, B. D. Rainford, and O. Vogt, Journal of Physics C-Solid State Physics **11**, 1173 (1978).
[16]    P. Fischer, B. Lebech, G. Meier, B. D. Rainford, and O. Vogt, Journal of Physics C-Solid State Physics **11**, 345 (1978).
[17]    B. Lebech, K. Clausen, and O. Vogt, Journal of Physics C-Solid State Physics **13**, 1725 (1980).
[18]    B. Halg and A. Furrer, Physical Review B **34**, 6258 (1986).
[19]    M. Kohgi, K. Iwasa, and T. Osakabe, Physica B-Condensed Matter **281**, 417 (2000).
[20]    K. Iwasa, A. Hannan, M. Kohgi, and T. Suzuki, Physical Review Letters **88**, 207201 (2002).
[21]    S. Jang, R. Kealhofer, C. John, S. Doyle, J. S. Hong, J. H. Shim, Q. Si, O. Erten, J. D. Denlinger, and J. G. Analytis, Science Advances **5**, eaat7158 (2019).
[22]    M. Escorne, A. Mauger, D. Ravot, and J. C. Achard, Journal of Physics C-Solid State Physics **14**, 1821 (1981).
[23]    L. D. Ye, T. Suzuki, C. R. Wicker, and J. G. Checkelsky, Physical Review B **97**, 081108 (2018).

[24]   J. M. Lock, A. B. Pippard, and D. Shoenberg, Proceedings of the Cambridge Philosophical Society **47**, 811 (1951).
[25]   C. A. Bryant and P. H. Keesom, Physical Review **123**, 491 (1961).
[26]   A. Scheie, Journal of Low Temperature Physics **193**, 60 (2018).
[27]   A. Abragam and B. Bleaney, *Electron paramagnetic resonance of transition ions* (Clarendon P., 1970).
[28]   D. P. Landau, B. E. Keen, B. Schneider, and W. P. Wolf, Physical Review B-Solid State **3**, 2310 (1971).
[29]   K. Niira, Physical Review **117**, 129 (1960).
[30]   A. I. Akhiezer, V. G. Bariakhtar, and M. I. Kaganov, Uspekhi Fizicheskikh Nauk **71**, 533 (1960).
[31]   C. Herring, Physical Review **52**, 0365 (1937).
[32]   C. K. Majumdar and D. K. Ghosh, Journal of Mathematical Physics **10**, 1388 (1969).
[33]   C. K. Majumdar and D. K. Ghosh, Journal of Mathematical Physics **10**, 1399 (1969).
[34]   A. Honecker, J. Schulenburg, and J. Richter, Journal of Physics-Condensed Matter **16**, S749, Pii s0953-8984(04)74155-7 (2004).
[35]   F. Mila, Physical Review Letters **81**, 2356 (1998).
[36]   K. Hida, Journal of the Physical Society of Japan **70**, 3673 (2001).
[37]   E. Gornik, R. Lassnig, G. Strasser, H. L. Stormer, A. C. Gossard, and W. Wiegmann, Physical Review Letters **54**, 1820 (1985).
[38]   X. Z. Yu, Y. Onose, N. Kanazawa, J. H. Park, J. H. Han, Y. Matsui, N. Nagaosa, and Y. Tokura, Nature **465**, 901 (2010).

Supplemental Information
To
Extremely Weakly Interacting $\Delta S_z = 0$ and $\Delta S_z = 1$ Excitations and Evidence for Fractional Quantization in a Magnetization Plateau: CeSb

P. G. LaBarre[1], B. Kuthanazhi[2,3], C. Abel[2,3], P. C. Canfield[2,3], and A. P. Ramirez[1]

[1] Department of Physics, University of California Santa Cruz, Santa Cruz, California 95064 USA
[2] Department of Physics and Astronomy Iowa State University, Ames, Iowa 50011, USA
[3] Ames Laboratory, Ames, Iowa, 50011, USA

Synthesis

The crystals of CeSb used in this study were grown from a high temperature melt utilizing a third element, in this case tin, as a flux. The elements Ce (Ames Laboratory, 99.99%), Sb (Aesar, 99.99%) and Sn (Aesar, 99.991%) were placed in an alumina crucible with atomic ratios of Ce:Sb:Sn = 6:6:88. The crucible was then sealed in a partial pressure of argon in a quartz ampoule. The melt was heated to 1150 C over 5 h, held at 1150 C for 5 h, then slowly cooled over a period of 200 h to 800 C, at which temperature the excess flux was decanted from the crystals.

Determination of the lattice contribution to C(T)

The lattice contribution to $C(T)$ in CeSb was determined using a single crystal of non-magnetic LaSb, grown for this purpose. The $C(T)/T$ data for LaSb at $H = 0.1T$, along with those of CeSb at $H = 0$, are shown in Fig. 1. The applied field used in the LaSb measurement was required to suppress an anomaly at $T = 3.7 K$, which was assumed to be due to the superconducting transition in residual Sn flux [1,2] used in the synthesis. No other $H$ dependence was observed. The size of this anomaly, about 9% of the total $C(T)$, implies that Sn constitutes a mass fraction of 7% in LaSb sample. Thus, for the LaSb *specific heat* we took the measured heat capacity data at 0.1T, subtracted the heat capacity of the derived mass of Sn, and normalized the result by the reduced mass of LaSb. The Sn correction constitutes 7.6% (6.1%) of the lattice and 1.6% (0.1%) of the total CeSb $C(T)$ at 3K (9K) respectively.

While the subtraction of LaSb specific heat accounts for the phonon contribution to the total specific heat CeSb, another lattice contribution is associated with the latent heat of the first order structural transition from the NaCl structure to a tetragonal structure. This transition occurs simultaneously with the Neél transition at 16K, and is partially indicated by the sharp peak seen in Fig. 1. We found that, as expected for a first-order transition, the height of the peak, and thus the integrated entropy, depends strongly on the size of the temperature step in the heat capacity measurement [3]. Thus, while our specific heat data below 15K agree with a previous measurement [4] we find a smaller value of integrated entropy through the Neél transition. While this observation has no relevance to the conclusions of the present work, if the structural latent heat is not accounted for accurately, it can be wrongly construed to imply that the Neél order involves substantial participation from the $\Gamma_8$ quartet states. As emphasized in the main text, however, the good agreement between our observed saturation moment and that expected for the $\Gamma_7$ manifold provides assurance that the physics described here is that of a Kramers doublet. This point is further emphasized below.

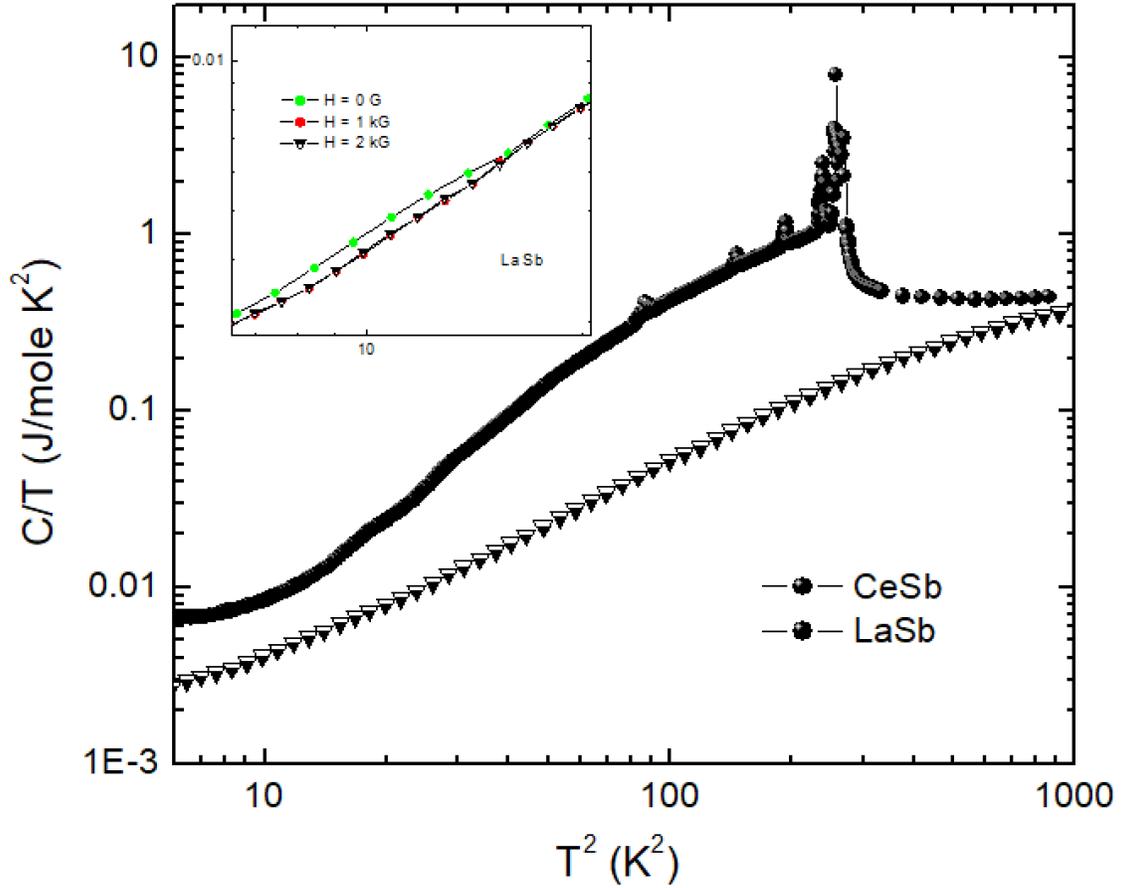

Figure 1: C/T v $T^2$ of CeSb at $H = 0$ and LaSb at $H = 0.2$ T. The sharp peaks seen in the CeSb data indicate the $H = 0$ occurrence of the multiple phases observed outside of the 1/3rd plateau region.

Specific heat of a free-spin $\Gamma_8$ quartet

In Fig. 2 are plotted data for $C(T)/T$ of a free-spin quartet state with S = 3/2, a Lande g-factor of 6/7, and a zero-field energy of $24k_B$, as appropriate for CeSb, in a field of 3T. Also shown is $C(T)/T$ for CeSb at 3T. As stated in the main text, we see that the specific heat of the quartet state is much larger than that that of CeSb. Clearly the density of states for these crystal field excitations is significantly reduced due to the collective effect of the spin-spin interaction within the plateau.

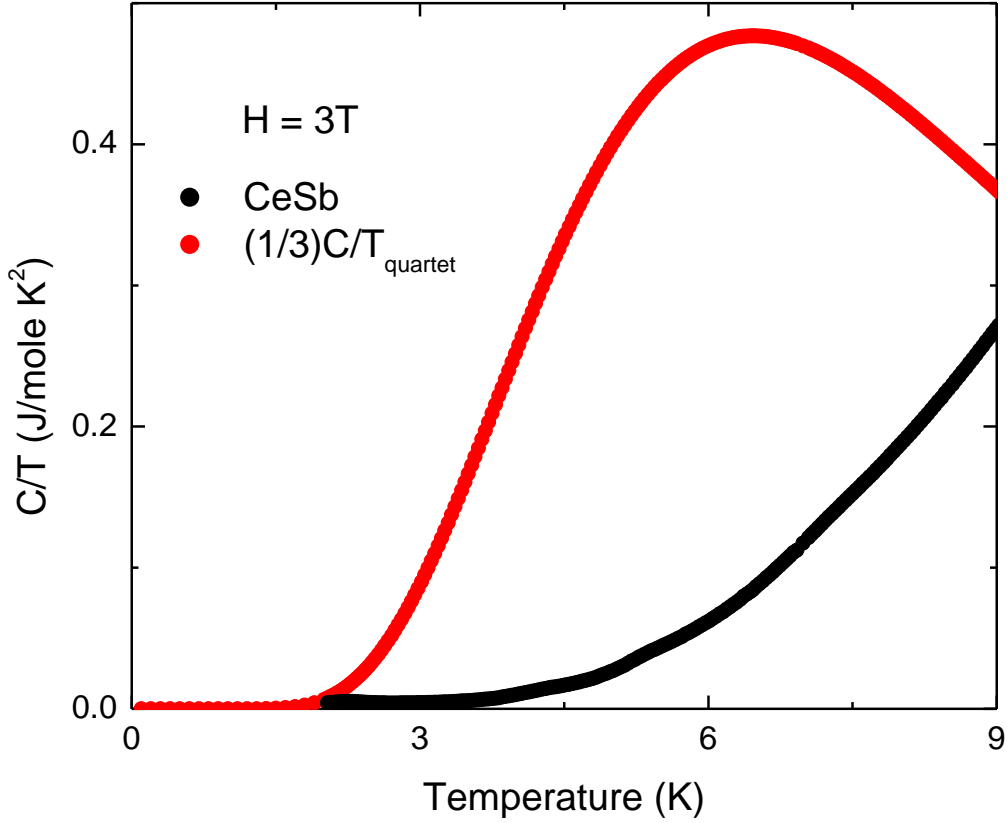

Figure 2: C/T v T of CeSb and 1/3$^{rd}$ of the specific heat due to a quartet state lifted by a crystal field gap of T = 24K in an external field of H = 3 T.

Intra-plateau *H-T* diagram

In Fig. 3 are plotted the *H-T* dependence of the mini-jumps seen within the 1/3$^{rd}$ plateau. As stated in the main text, the precise positions acquired from $dM/dH$ show that the central, and largest, of the jumps asymptotically approaches as $T \to 2K$, a field value corresponding to $H = H_l + f_p \Delta H_{1/3}$ where $\Delta H_{1/3} = 1.7 \pm 0.1$ T is the width of the 1/3$^{rd}$ plateau and $f_p = 1/2$. The plateau width is determined at each temperature as small variations in the width, lower, and upper field boundaries are seen from $2 - 9\ K$, the adjusted $f_p = 1/2$ field value at each temperature is shown in open circles of fig. 4 (a) of the main text.

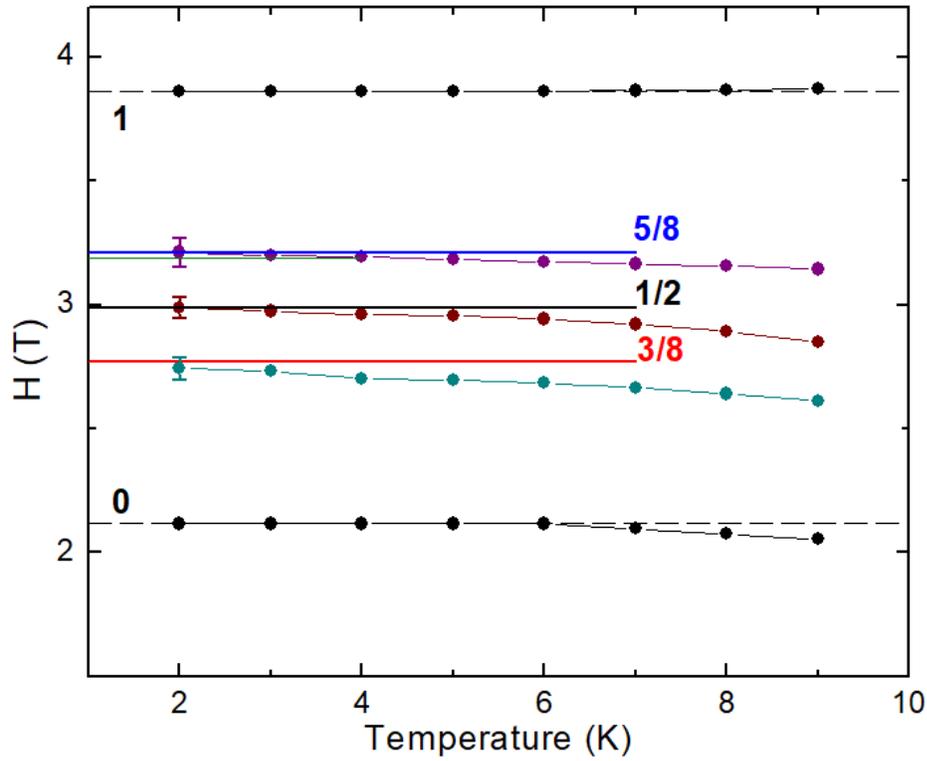

Figure 3: Positions of the mini jumps seen in the 1/3rd plateau region of CeSb in the *H-T* plane. The red, black and blue lines correspond to fractional field values relative to the field width of the 1/3 plateau at 2K.

References

[1]     J. M. Lock, A. B. Pippard, and D. Shoenberg, Proceedings of the Cambridge Philosophical Society **47**, 811 (1951).
[2]     C. A. Bryant and P. H. Keesom, Physical Review **123**, 491 (1961).
[3]     A. Scheie, Journal of Low Temperature Physics **193**, 60 (2018).
[4]     J. Rossat-Mignod, P. Burlet, H. Bartholin, O. Vogt, and R. Lagnier, Journal of Physics C-Solid State Physics **13**, 6381 (1980).